\documentclass[a4paper]{jpconf}
\usepackage{graphicx}
\begin{document}
\title{Superfluid Phase Stability of $^3$He in Axially Anisotropic Aerogel}

\author{J. Pollanen, J.P. Davis, B. Reddy, K.R. Shirer, H. Choi and W.P. Halperin}

\address{Department of Physics and Astronomy, Northwestern University, Evanston, IL 60208, USA}

\ead{j-pollanen@northwestern.edu}

\begin{abstract}
Measurements of superfluid $^3$He in $98\%$ aerogel demonstrate the existence of a metastable \emph{A}-like phase and a stable \emph{B}-like phase.  It has been suggested that the relative stability of these two phases is controlled by anisotropic quasiparticle scattering in the aerogel.  Anisotropic scattering produced by axial compression of the aerogel has been predicted to stabilize the axial state of superfluid $^3$He.  To explore this possiblity, we used transverse acoustic impedance to map out the phase diagram of superfluid $^3$He in a $\sim 98$\% porous silica aerogel subjected to 17\% axial compression.  We have previously shown that axial anisotropy in aerogel leads to optical birefringence and that optical cross-polarization studies can be used to characterize such anisotropy.  Consequently, we have performed optical cross-polarization experiments to verify the presence and uniformity of the axial anisotropy in our aerogel sample.  We find that uniform axial anisotropy introduced by 17\% compression does not stabilize the \emph{A}-like phase.  We also find an increase in the supercooling of the \emph{A}-like phase at lower pressure, indicating a modification to \emph{B}-like phase nucleation in \emph{globally} anisotropic aerogels. 
\end{abstract}

\section{Introduction}
High porosity silica aerogels are the only known means for studying the effect of impurities on the otherwise completely pure superfluid phases of $^{3}$He.  Recently there has been substantial interest in exploring the role of structural anisotropy in the aerogel and how it pertains to superfluid phase stability, textural orientation, spin dynamics, and nucleation.  The narrow window of \emph{A}-like phase stability might be explained by the existence of anisotropy in the aerogel structure \cite{Thu98,Vic05,Aoy06}.  The possibility of using anisotropic aerogels to stabilize superfluid phases not present in bulk $^{3}$He, such as the polar phase, has also been explored theoretically \cite{Aoy06} and experimentally \cite{Dav06,Elb08,Dav08}.  Experimental work has shown that anisotropy affects the orientational degrees of freedom in both the \emph{A}-like and \emph{B}-like phases \cite{Kun07,Dmi07} and anisotropic aerogels allow for the realization of novel spin dynamical states not possible in bulk \cite{Sat08}.  The nucleation of the \emph{B}-like phase may also be modified by anisotropy in the aerogel \cite{Dav08,Kad08}.  We report on measurements of the transverse acoustic impedance of superfluid $^{3}$He confined in a 97.8\% porous aerogel with well defined axial anisotropy to study the role of anisotropy on phase stability and nucleation.      

\section{Axial Strain and Optical Birefringence}
Optical birefringence in a transparent or translucent material results from an anisotropic dielectric constant, which leads to an optical axis in the material.  The optical axis acts as a linear polarizer whose direction and distribution throughout the sample can be determined by viewing the material between crossed polarizers \cite{Hec98}.     

\begin{figure}[t]
\centerline{\includegraphics[height=0.07\textheight]{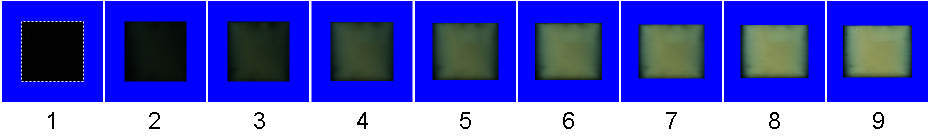}}
\caption{\label{fig1}(Color online) Optical birefringence of a $98\%$ porosity aerogel, initially isotropic before it was subjected to increasing axial strain.  The strain increases from left to right: 1 (unstrained), 2 (2.3\% strain), 3 (4.7\%), 4 (7.0\%), 5 (9.3\%), 6 (11.6\%), 7 (14.0\%), 8 (16.3\%), 9 (18.6\%).}
\end{figure}

We have previously shown that axial strain can be used to produce \emph{global} structural anisotropy in high porosity silica aerogel and that this anisotropy can be characterized by optical birefringence \cite{Pol08}.  Fig.~\ref{fig1} depicts how axial strain introduces anisotropy into an initially isotropic 98\% porous aerogel.  All aerogels discussed here were synthesized at Northwestern University via a one-step sol-gel process followed by supercritical drying \cite{Pol08}.  

\begin{figure}[h]
\centerline{\includegraphics[height=0.07\textheight]{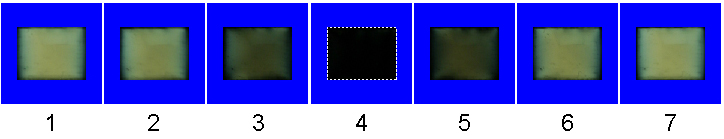}}
\caption{\label{fig2}(Color online) Optical birefringence of a $98\%$ porosity aerogel strained by 18.6\%.  The labels are associated with the rotation of the polarizers relative to the cylinder axis: 1 (45$^\circ$, 135$^\circ$), 2 (60$^\circ$, 150$^\circ$), 3 (75$^\circ$, 165$^\circ$), 4 (90$^\circ$, 180$^\circ$), 5 (105$^\circ$, 195$^\circ$), 6 (120$^\circ$, 210$^\circ$), 7 (135$^\circ$, 225$^\circ$).}
\end{figure}

A cylindrical aerogel sample was placed between two crossed polarizers and illuminated with diffuse white light.  The sample was oriented vertically and images were captured with a digital camera.  The polarizers were oriented with one at 45$^\circ$ and the other at 135$^\circ$ relative to the cylinder axis.  Panel 1 in Fig.~\ref{fig1} shows that the unstrained gel does not impose any preferred direction of polarization when placed between crossed polarizers.  We found this to be the case for all orientations of the polarizers with respect to the cylinder axis.  We also found that light propagating along the cylinder axis was not transmitted before the sample was strained .  Consequently, we refer to the nature of this aerogel sample prior to the application of strain as homogeneously isotropic.  The increasing intensity of light transmitted through the sample with increasing strain in Panels 2-9 indicate how global anisotropy was introduced as the sample was strained up to 18.6\%.
To determine the direction of the optical axis in the aerogel, we rotated the polarizers while keeping them crossed.  The rotation sequence for the sample at 18.6\% strain is presented in Fig.~\ref{fig2}.  Intensity maxima (minima), seen when the polarizers are at 45$^\circ$ and 135$^\circ$ (90$^\circ$ and 180$^\circ$), are consistent with the optical axis being oriented along the strain axis.

\section{Acoustic Impedance Experiment}
Transverse acoustic impedance is sensitive to phase transitions of superfluid $^{3}$He in aerogel, and the technique has been used to map out the pressure-temperature phase diagram in unstrained aerogel samples \cite{Ger01,Ger02}.  To explore the role of axial strain, we have performed measurements of transverse acoustic impedance at 17.6 MHz in a sample of 97.8\% porous aerogel grown onto the surface of an \emph{AC}-cut quartz piezoelectric transducer and strained by 17\% \cite{Dav08}.  The impedance was measured using a continuous wave RF-bridge \cite{DaT08,Ham89}.  The sample cell was cooled in liquid $^{3}$He using a dilution refrigerator, followed by adiabatic nuclear demagnetization.  We monitored the temperature of the $^{3}$He by measuring the temperature dependent magnetic susceptibiliy of a paramagnetic salt (LCMN) \cite{Ham89}, calibrated with respect to the Greywall temperature scale \cite{Gre86}.

\begin{figure}[h]
\centerline{\includegraphics[height=0.175\textheight]{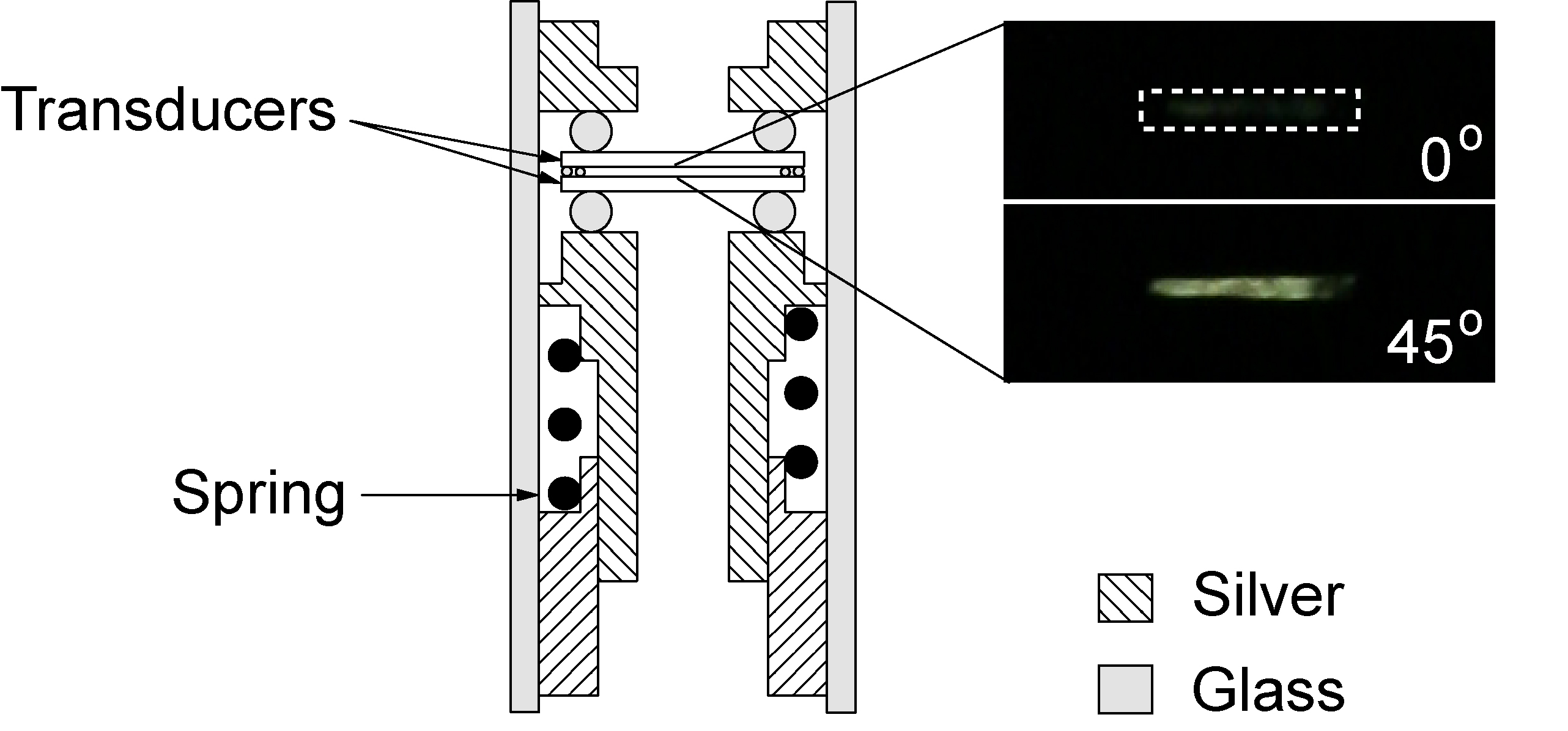}}
\caption{\label{fig3}(Color online) Aerogel acoustic compression cell.  Aerogel was grown directly between, and onto, the two quartz transducers and compressed {\it in situ}.  The inset depicts optical birefringence of the aerogel with the polarizers oriented at (0$^\circ$, 90$^\circ$) (top panel) and (45$^\circ$, 135$^\circ$) (lower panel) relative to the surface normal of the transducer.}
\end{figure}

The sample cell consisted of an open cavity between two parallel transducers separated by two spacer wires of diameter 0.0305 cm.  Alongside and between these spacer wires were two smaller wires of diameter 0.0254 cm.  In this configuration the cavity was held under tension by a stainless steel spring, see Fig.~\ref{fig3}.  Aerogel was grown directly in the cavity and onto the surfaces of the transducers.  Excess aerogel around the cavity was removed such that the outer surfaces of the transducers would be in contact with bulk $^{3}$He.  Then the large diameter spacers wires were removed and the spring compressed the aerogel onto the small diameter wires.  This procedure resulted in 17\% strain.  To verify the existence of global anisotropy in our sample after the compression procedure, we performed optical birefringence measurements on the aerogel cavity after the experiment was completed.  Fig.~\ref{fig3} depicts a diagram of the sample cell and the inset shows the birefringence of the aerogel in the acoustic cavity with the polarizers oriented at (0$^\circ$, 90$^\circ$) (top panel) and (45$^\circ$, 135$^\circ$) (lower panel) relative to the surface normal of the transducer.  Fig.~\ref{fig3} clearly indicates that the compression cell introduces global anisotropy in our aerogel sample and that the magnitude and direction of the anisotropy are consistent with our previous work on compressed aerogels \cite{Pol08}. 

\section{Results and Discussion}
\begin{figure}
\centerline{\includegraphics[height=0.25\textheight]{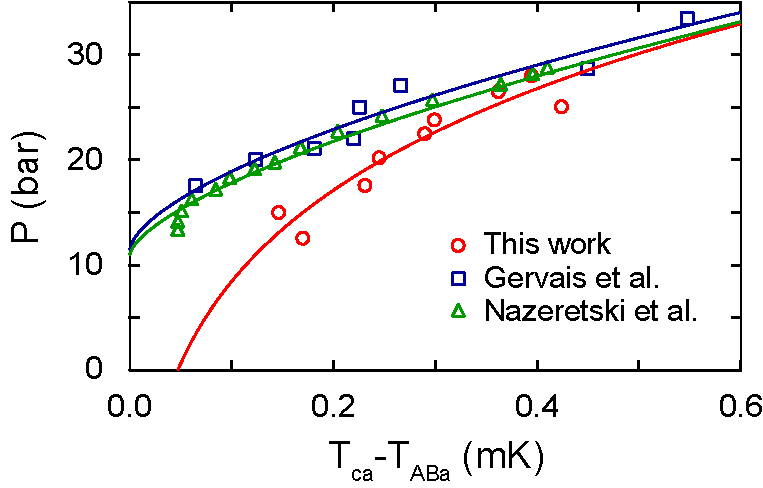}}
\caption{\label{fig4}(Color online) Increased supercooling of the \emph{A}-like to \emph{B}-like transition as a function of pressure for a 17\% axially compressed aerogel (open red circles) along with the previous results of Gervais {\it{et. al}} \cite{Ger01,Ger02} (open green squares) and Nazaretski {\it{et. al}} \cite{Naz04} (open blue triangles).  Curves are guides to the eye.}
\end{figure}

On cooling from the normal state we saw a clear indication of the aerogel-superfluid transition, \emph{T}$_{ca}$ and the \emph{A}-like to \emph{B}-like transition, \emph{T}$_{ABa}$.  However, on warming the sample from the \emph{B}-like phase we did not observe any other transitions until we reached \emph{T}$_{ca}$ \cite{Dav08}. Our tracking experiments \cite{Dav08} near \emph{T}$_{ca}$ reveal that the window of \emph{A}-like phase stability is identical to previous studies \cite{Vic05,Ger02}.   Therefore, we conclude that 17\% axial strain does not stabilize the \emph{A}-like phase.  In Fig.~\ref{fig4} we present results on the amount of supercooling in our axially compressed aerogel. For comparison we include the results of Gervais \emph{et al.} \cite{Ger01,Ger02} and Nazaretski \emph{et al.} \cite{Naz04} on presumably isotropic aerogel samples.  We find that there was a pronounced increase in supercooling of the \emph{A}-like phase at pressures below 20 bar for our sample.  This indicates that the nucleation mechanism for the \emph{B}-like phase is suppressed by the axial anisotropy of our sample.  This implies that the metastability of the \emph{A}-like is enhanced by the presence of anisotropic scattering inside the aerogel.      

\section{Conclusions}
Our results show that global anisotropy produced by 17\% axial anisotropy inhibits the nucleation of the \emph{B}-like phase of superfluid $^{3}$He in aerogel but does not modify the thermodynamic stability of the \emph{A}-like phase.

\ack
We acknowledge support from the National Science Foundation, DMR-0703656.  The authors would like to thank J.A. Sauls for valuable theoretical insights.  We are indebted to P.E. Wolf and F. Bonnet for introducing us to the technique of optical polarization studies for aerogel characterization.


\section{References}
\begin{thebibliography}{9}
\bibitem{Thu98} E.V. Thuneberg, S.K. Yip, M. Fogelstr\"{o}m and J.A. Sauls, \emph{Phys. Rev. Lett.} {\bf80}, 2861 (1998).
\bibitem{Vic05} C.L. Vicente, H.C. Choi, J.S. Xia, W.P. Halperin, N. Mulders and Y. Lee, \emph{Phys. Rev. B} {\bf72}, 094519 2005.
\bibitem{Aoy06} K. Aoyama and R. Ikeda. \emph{Phys. Rev B} {\bf73}, 060504 (2006).
\bibitem{Dav06} J.P. Davis, H. Choi, J. Pollanen and W.P. Halperin, \emph{AIP Conf. Proc.} {\bf850}, 239 (2006).
\bibitem{Elb08} J. Elbs, Yu.M. Bunkov, E. Collin, and H. Godfrin. \emph{Phys. Rev. Lett.} {\bf100}, 215304 (2008).
\bibitem{Dav08} J.P. Davis, J. Pollanen, B. Reddy, K.R. Shirer, H. Choi and W.P. Halperin, \emph{Phys. Rev. B} {\bf77}, 140502(R) (2008).
\bibitem{Kun07} T. Kunimatsu, T. Sato, K. Izumina, A. Matsubara, Y. Sasaki, M. Kubota, O. Ishikawa, T. Mizusaki and Yu. M. Bunkov, \emph{JETP Lett.} {\bf86}, 216 (2007).
\bibitem{Dmi07} V.V. Dmitriev, D.A. Krasnikhin, N. Mulders, V.V. Zavjalov, and D.E. Zmeev. \emph{Pis'ma v ZhETF} {\bf86} (2007).
\bibitem{Sat08} T. Sato, T. Kunimatsu, K. Izumina, A. Matsubara, M. Kubota, T. Mizusaki and Yu. M. Bunkov, arXiv:0804.2994v1 (2008).
\bibitem{Kad08} R. Kado, H. Nakagawa, K. Obara, H. Yano, O. Ishikawa, T. Hata. \emph{J. Low Temp. Phys.} {\bf150} (2008). 
\bibitem{Hec98} E. Hecht, {\em Optics 3rd ed.}, 330, (Addison-Wesley, Massachusetts) (1998).
\bibitem{Pol08} J. Pollanen, K. Shirer, S. Blinstein, J.P. Davis, H. Choi, T.M. Lippman, L.B. Lurio, and W.P. Halperin, accepted for publication in \emph{J. Non-Crystalline Solids} (2008).
\bibitem{Ger01} G. Gervais, T.M. Haard, R. Nomura, N. Mulders and W.P. Halperin, \emph{Phys. Rev. Lett.} {\bf87}, 035701 (2001).
\bibitem{Ger02} G. Gervais, K. Yawata, N. Mulders and W.P. Halperin, \emph{Phys. Rev. B} {\bf66}, 054528 (2002).
\bibitem{DaT08} J.P. Davis, Ph.D. thesis, Northwestern University, (unpublished) (2008).
\bibitem{Ham89} P.J. Hamot, H.H. Hensley and W.P. Halperin, \emph{J. of Low Temp. Phys.} {\bf77}, 429 (1989).
\bibitem{Gre86} D.S. Greywall, \emph{Phys. Rev. B} {\bf33}, 7520 (1986).
\bibitem{Naz04} E. Nazaretski, N. Mulders and J.M. Parpia, \emph{JETP Lett.} {\bf79}, 383 (2004).
\end{thebibliography}
\end{document}